\documentstyle[12pt,a4,amssymb,times]{article}

\def\vphi{\varphi}
\def\D{\Delta}
\def\e{\epsilon}
\newcommand{\be}{\begin{equation}}
\newcommand{\ee}{\end{equation}} 
\newcommand{\bea}{\begin{eqnarray}}
\newcommand{\eea}{\end{eqnarray}}

\begin{document}

\begin{titlepage}

\begin{flushright} 
{\tt 	FTUV/98-21\\ 
	IFIC/98-21\\ }
\end{flushright}

\bigskip

\begin{center}

{\bf{\Large
    Free field realization of \\
cylindrically symmetric Einstein gravity
\footnote{Work partially supported by the 
{\it Comisi\'on Interministerial de Ciencia y Tecnolog\'{\i}a}\/ 
and {\it DGICYT}}}}

\bigskip 

 J.~Cruz \footnote{E-mail address: cruz@lie.uv.es},
 A. Mikovi\'c \footnote{E-mail address: mikovic@lie.uv.es. On leave of absence 
from Institute of Physics, P.O.Box 57, 11001 Belgrade, Yugoslavia}
and J.~Navarro-Salas \footnote{E-mail address: jnavarro@lie.uv.es}

\end{center}

\begin{center}

\footnotesize
	 Departamento de F\'{\i}sica Te\'orica and 
	IFIC, Centro Mixto Universidad de Valencia-CSIC.
	Facultad de F\'{\i}sica, Universidad de Valencia,	
        Burjassot-46100, Valencia, Spain 
\end{center}

\normalsize 

\bigskip 
\bigskip
\begin{center}
			{\bf Abstract}
\end{center}	
Cylindrically reduced  Einstein gravity can be regarded as an 
$SL(2,R)/SO(2)$ sigma model coupled to 2D dilaton gravity. By 
using the corresponding 2D diffeomorphism algebra of constraints
and the asymptotic behaviour of the Ernst equation we show that
 the theory can be mapped by a canonical transformation
into a set of free fields with a Minkowskian target space. We briefly
discuss the quantization in terms of these free-field variables, which is 
considerably simpler than in the other approaches.

\noindent
\bigskip
\bigskip
 
 PACS number(s): 04.60.Kz, 04.60.Ds
\end{titlepage}
 \newpage

Two Killing vector reductions of 4D Einstein equations are exactly integrable
2D models \cite{bz,m,KS,ah,bj}, and therefore
offer an interesting arena to investigate the quantization 
of the gravitational field \cite{K,A,nk}.
These models are described  by a 4D line-element of the form
\be
ds^{2}=g_{\mu\nu}(x)dx^{\mu}dx^{\nu}+\vphi (x)\Lambda_{ab} (x) 
d\chi^a d\chi^b \quad,
\ee
where $x^{\mu}$ are the 2D coordinates,  $\partial /\partial \chi^a$ are
the Killing vectors and $\det\Lambda =1$.  
The corresponding Einstein equations can be derived from an action
for a 2D dilaton gravity coupled to an $SL(2,R)/SO(2)$ coset
space $\sigma$-model \cite{bmg}
\be
S=\int d^2x\sqrt{-g} \vphi\left[R-{1\over4}tr(\Lambda^{-1}\nabla^{\mu}\Lambda
\Lambda^{-1}\nabla_{\mu}\Lambda)\right] 
\>.\label{ai}
\ee
Depending on the choice of the coordinates and the spatial topology one can 
describe cylindrical gravitational waves  
$\left(x^\mu =(t,r),\ \chi^a =\left(z,\phi\right)\right)$, axisymmetric 
stationary solutions
$\left(x^\mu =(z,r),\ \chi^a =\left(t,\phi\right)\right)$ or Gowdy cosmologies
$(x^\mu =(t,x),\ \chi^a =(y,z)$ $x,y,z \in S^1$) where $t$ is a time 
coordinate while $r,\phi,z$ are cylindrical coordinates and $x,y,z$ are 
Cartesian coordinates. 

The usual approach to study (\ref{ai}) is to work in the reduced phase space 
formalism ($\vphi = r$ gauge for cylindrical waves), so that the complete 
dynamics is contained in the Ernst equation for the matrix $\Lambda$
\be
\nabla_{\mu}(\vphi \Lambda^{-1}\nabla^{\mu}\Lambda)=0
\>.\label{ern}
\ee
Note that (\ref{ern}) has a duality symmetry, so that if 
\be
\Lambda=\frac1{\D} \left(\begin{array}{cc}h^2 + \D^2 & h\\h&1
\end{array}\right)
\> \label{p1}
\ee
is a solution then  
\be
\tilde\Lambda=\frac{\D}{r} \left(\begin{array}{cc}h^2 +\frac{r^2}{\D^2} & 
\tilde h\\ \tilde h &1 \end{array}\right)
\>\label{p2}
\ee
is also a solution, provided that $^* d \tilde h = \frac{r}{\D^2} dh$. This 
symmetry implies that $d{\tilde s}\sp 2$
can have the asymptotic behaviour of a flat metric in cylindrical coordinates,
i.e. 
$d{\tilde s}\sp 2 \sim -dt\sp 2 + dr\sp 2 + r\sp 2 d\phi\sp 2 + dz\sp 2$,
which is relevant for the case of cylindrical waves, which we are going to
study in detail.

The aim of this letter is to show that there is a considerable advantage if
one does not fix the gauge completely, and consequently uses the
special properties of the 2D diffeomorphism algebra of the constraints.
The special role of 2D diffeomorphisms has been already recognized in 
\cite{nk},
although in that work the reduced phase space formalism and an
abelianized form of the constraints was used. In this paper we employ
the full constraint structure, which generalizes
the technique used for the case of cylindrical gravitational waves with one
polarization \cite{Cruz}, when the matrix $\Lambda$ takes the diagonal form 
\be
\Lambda=\left(\begin{array}{cc}e^{f}&0\\0&e^{-f}
\end{array}\right)\>.\label{abe} 
\ee
In that case it is possible to show that a canonical transformation exists 
which maps the constraints into a free-field form. We will show that the same 
can be done in the case of two polarizations, so that the theory is mapped into
a theory of four free fields.

This looks as a surprising result, but there are general arguments which 
suggest
that this is possible.
 The Poisson bracket algebra of
the constraints for (\ref{ai}) is the 2D diffeomorphism
algebra. This algebra admits  representations quadratic in canonical variables
\be G_0 = \frac12 \left( \eta^{ij} P_i P_j + 
\eta_{ij} Q^{i\prime } Q^{j\prime} \right) \quad,\quad 
G_1 = P_i Q^{i\prime } \quad,\label{rep}\ee
where $G_0$ is the hamiltonian constraint and $G_1$ is the 
spatial diffeomorphism constraint. The
prime denotes differentiation with respect to the spatial coordinate 
($r$ in our case),
$i,j=1,...,n$ and $\eta_{ij}$ is a flat Minkowskian metric.
 The quadratic
representations are not
 possible in higher dimensions, since then the constraint algebra has 
structure functions which are not constant. The representation (\ref{rep})
implies that $Q^i$ are free fields and 
since (\ref{ai}) is an integrable 2D theory
one then expects to find a canonical  transformation from the initial
canonical
variables to $(P_i,Q^i)$ variables. 
This argument also explains why it was possible to find free field canonical 
variables
for many examples of integrable 2D dilaton gravity theories
\cite{CJZ,Cruz}.
Also note that  by writing $g_{\mu\nu}=e^{2\rho}\hat g_{\mu\nu}$, 
where $\hat g_{\mu\nu}$ is a fixed background metric, 
the action (\ref{ai}) becomes 
a nonlinear $\sigma$-model with a four-dimensional target space
\be
S=\int d^2 x\sqrt{-\hat g}
\left[G_{ij}(X)\hat\nabla_{\mu}X^i\hat\nabla^{\mu}X^j+ \Phi(X) \hat R\right]
\>.\label{sigma}
\ee
The general covariance of (\ref{ai}), or equivalently the background 
independence of (\ref{sigma}),
implies that the theory is 
a conformally invariant field theory.
In fact, the couplings $G_{ij}(X)$ and $\Phi(X)$ satisfy the lowest order 
$\beta$-functions equations
in the standard loop expansion for dilaton-gravity.
Since in the abelian case this conformal field theory is a free-field theory,
one also expects that this will happen in the non-abelian case. 
 In the
abelian case (\ref{abe}) the Ernst equation reduces to the cylindrical 
Laplace equation and
the theory can be explicitly solved. Furthermore,
the asymptotic behaviour of the solutions allows one to show that the theory 
is equivalent to a theory of three free fields
with a Minkowskian target-space.
Although the Ernst equation cannot be explicitly solved in the general case,
the asymptotic behaviour of the solutions in the weak coupling regime 
$\vphi\rightarrow\infty$
is enough to show that the underlying conformal field theory  
is still described by a set of free fields.

The SL(2,R)/SO(2) coset space can be parametrized in the following way
\be
\Lambda=\left(\begin{array}{cc}e^{f}+h^2e^{-f}&e^{-f}h\\e^{-f}h&e^{-f}
\end{array}\right)
\>.
\ee
This is the parametrization (\ref{p1}) with $\D = e^f$. The action 
(\ref{ai}) then becomes
\be
S=\int d^2x\sqrt{-g}\vphi\left[R-{1\over2}\left(\nabla f\right)^2-
{1\over2}e^{-2f}
\left(\nabla h\right)^2\right] \>.\label{aiii}
\ee
The corresponding equations of motion in the conformal gauge are given by
\be
\partial_+\partial_-\vphi=0\>,\label{aiv}
\ee
\be
4\partial_+\partial_-\rho+\partial_+f\partial_-f+
e^{-2f}\partial_+h\partial_-h=0\>,\label{av}
\ee
\be  
\partial_+(\vphi\partial_-f)+\partial_-(\vphi\partial_+f)+
2\vphi e^{-2f}\partial_+h\partial_-h=0\>,
\label{avi}
\ee
\be
\partial_+(\vphi e^{-2f}\partial_-h)+\partial_-(\vphi e^{-2f}\partial_+h)
=0\>,\label{avii}
\ee
\be
C_{\pm}=  2\partial_{\pm}^2\vphi - 4\partial_{\pm}\vphi
\partial_{\pm}\rho + \vphi\left[\left(\partial_{\pm}f\right)^2+e^{-2f}(
\partial_{\pm}h)^2\right]=0     \>,
\ee
where $ C_\pm = \frac12 (G_0 \pm G_1 )$ are the constraint equations.
The free field equation (\ref{aiv}) has the obvious solution
\be
\vphi=A_+(x^+)+A_-(x^-)
\>,
\ee
with $A_{\pm}$ two arbitrary chiral functions,
while the conformal factor $\rho$ can be expressed as
\be
\rho=a_+(x^+)+a_-(x^-)+{1\over4}\int_{x^+}^{\infty}dy^+ \int^{x^-}_{-\infty}
dy^- \left[\partial_+f\partial_-f
+e^{-2f}\partial_+h\partial_-h\right]
\>,
\ee
where $a_{\pm}$ are other two arbitrary chiral functions.
By inserting the last two expressions into the constraints, we obtain 
\bea
C_{\pm}&=& 2\partial_{\pm}^2A_{\pm} - 4\partial_{\pm}A_{\pm}\partial_{\pm}
a_{\pm}\nonumber\\&&+ \partial_{\pm}A_{\pm}\int^{x^{\mp}}_
{\mp\infty} dy^\mp \left[\partial_+f\partial_-f
+e^{-2f}\partial_+h\partial_-h\right]\nonumber\\&&
+
\vphi\left[\left(\partial_{\pm}f\right)^2+e^{-2f}(\partial_{\pm}h)^2\right]=0\>.
\eea
The Bianchi identities $\partial_\mp C_\pm = 0$ imply
that $\partial_{\mp}P_{\pm}=0$, where
\bea
P_{\pm}&=&
{1\over2}\partial_{\pm}A_{\pm}\int^{x^{\mp}}_{\mp\infty} dy^\mp
\left[\partial_+f\partial_-f
+e^{-2f}\partial_+h\partial_-h\right]\nonumber\\&&
+{1\over2}
\vphi\left[\left(\partial_{\pm}f\right)^2+e^{-2f}(\partial_{\pm}h)^2\right]
\>.\label{ep}
\eea
$P_{\pm}(x^{\pm})$ can be evaluated by taking the limits
$x^{\mp}\rightarrow \mp\infty$, since then the integral terms vanish, and hence
\be
P_{\pm}=\lim_{x^{\mp}\rightarrow \mp\infty} {1\over2}\vphi
\left[(\partial_{\pm}f)^2+e^{-2f}(\partial_{\pm}h)^2\right]\>.\label{p}
\ee
The main difference with respect to the abelian case
($h=0$) is that the equations
(\ref{avi}) and (\ref{avii}) for the fields $f$ and $h$ 
can not be solved explicitely.
However, for our purposes it is sufficient to know the asymptotic
behaviour of the solutions and  this can be found without
solving the equations.

If we
perform a change of variables $f\sqrt{\vphi}= \tilde F,\ 
h\sqrt{\vphi}=\tilde H$
where $\tilde F, \tilde H$ and their derivatives are bounded in the limit
$\vphi\rightarrow\infty$ and $A\left(x^+\right)$ and $B\left(x^-\right)$ are 
monotonic increasing (decreasing) functions
which go as $x^+(-x^-)$ when $x^{\pm}\rightarrow\infty\ (x^-\rightarrow -\infty)$,
then the equations (\ref{avi}) and (\ref{avii}) can be written as
\be
\partial_{+}\partial_{-}\tilde F +O\left(1\over\sqrt{\vphi}\right)=0\>,
\ee
\be
\partial_{+}\partial_{-} \tilde H +O\left(1\over\sqrt{\vphi}\right)=0\>.
\ee
Therefore in the limit $\vphi\rightarrow\infty$, one has
$f\sim {F\over\sqrt{A_+ + A_-}}$ and $h\sim {H\over\sqrt{A_++A_-}}$,
where $F$ and $H$ are bounded free fields with bounded derivatives.
In the abelian case this is explicitly realized because the
exact solution for the field $f$, 
\bea
f&=&{1\over2}\int_{-\infty}^{\infty} d\lambda\ 
J_0\left(
{\lambda\over2}(A_++A_-)\right)
\left[B_+(\lambda)e^{i{\lambda\over2}(A_+-A_-)}\right.\nonumber\\&&
\left.
+B_-(\lambda)
e^{-i{\lambda\over2}(A_+-A_-)}\right]
\>,
\eea
where $B_{\pm}(\lambda)$ are arbitrary coefficients,
behaves as 
\bea
f&\sim & {1\over\sqrt{(A_++A_-)}}\int_{-\infty}^{\infty}d\lambda
(\pi\lambda)^{-{1\over2}}\left[ B_+(\lambda)e^{i\lambda A_+}
e^{-i{\pi\over4}}+B_-(\lambda)e^{-i\lambda A_+}e^{i{\pi\over4}}\right.
\nonumber\\ &&
\left.
+B_+(\lambda)e^{-i\lambda A_-}e^{-i{\pi\over4}}+B_-(\lambda)e^{i\lambda A_-}
e^{i{\pi\over4}}\right]
\>,
\eea
when $A_++A_-\rightarrow\infty$ 
and therefore $F$ is a bounded free field and $\partial_+ F,
\partial_- F,...$
are also bounded.
By taking into account the asymptotic behaviour of $f$ and $h$, we can obtain from (\ref{p})
\be
P_{\pm}={1\over2}(\partial_{\pm}F)^2 +{1\over2}(\partial_{\pm}H)^2\>.
\ee
If one defines
\be
X^{\pm}=A_{\pm},\qquad \Pi_{\pm}=-4\partial_{\pm}a_{\pm}+2{\partial_
{\pm}^2A_{\pm}\over\partial_{\pm} A_{\pm}}
\>,\label{ct1}
\ee
the constraints take a free-field form
\be
C_{\pm}=\Pi_{\pm}\partial_{\pm}X^{\pm}+
\left(\partial_{\pm}F\right)^2+(\partial_{\pm}H)^2
\>.\label{ff}
\ee
In terms of the canonical variables, (\ref{ff}) can be written as
\be
C_{\pm}=\pm\Pi_{\pm} X^{\prime\pm}+
\frac14\left(\Pi_F \pm  F^{\prime} \right)^2+ \frac14\left(\Pi_H \pm H^{\prime} \right)^2
\>.\label{cff}
\ee
By performing a canonical transformation
\be 
2X^{\pm\prime} = \mp (\Pi_1 - \Pi_0) - X^{0\prime} - X^{1\prime} \quad,
\quad 2\Pi^{\pm} = -\Pi_0 - \Pi_1 \mp ( X^{1\prime} - X^{0\prime}) \, \label{ct}
\ee
the constraints take the form (\ref{rep}).

We can also find the exact expressions for the  free fields $F$ and $H$ in
terms of the initial variables. In order to do this, we split the 
expressions (\ref{ep}) as 
$P_{\pm}=P_{\pm}^{1}+P^{2}_{\pm}$ where
\be
P^{1}_{\pm}=
{1\over2}\vphi\left(\partial_{\pm}f\right)^2 +{1\over2}
\partial_{\pm}\vphi\int_{\mp\infty}^{x^{\mp}}dy^{\mp}\partial_+f
\partial_-f
+\int_{\mp\infty}^{x^{\mp}}dy^{\mp}\vphi e^{-2f}
\partial_{\pm}f\partial_+h\partial_-h
\>,\label{f2}
\ee
\be
P^2_{\pm}=
{1\over2}\vphi e^{-2f}(\partial_{\pm}h)^2 + {1\over2}\partial_{\pm}\vphi
\int^{x^{\mp}}_{\mp\infty}dy^{\mp}
e^{-2f}\partial_+h\partial_-h - 
\int^{x^{\mp}}_{\mp\infty}dy^{\mp}\vphi  e^{-2f}
\partial_{\pm}f\partial_+h\partial_-h
\>.\label{h2}
\ee
It is easy to check that the equations of motion imply that
\be
\partial_{\mp}P^{1}_{\pm}=\partial_{\mp}P^{2}_{\pm}=0
\>.
\ee
Thus we have divided $P_{\pm}$ into two free field contributions, and
it is clear that we can write 
\be  
P^{1}_{\pm}={1\over2}(\partial_{\pm}F)^2
\quad,\quad
P^{2}_{\pm}={1\over2}(\partial_{\pm}H)^{2}
\>.\label{ct2}
\ee
These two equations can serve as the defining relations for the 
free fields $F$ and $H$ in terms of the
initial variables. 

Therefore we have constructed a transformation leading to 
constraints quadratic in chiral variables and this implies that the
transformation is canonical \cite{Cruz}, since there is no other expression
for the symplectic form that together with (\ref{cff}) reproduces
 the Hamiltonian equations of motion for the
free fields $\Pi_\pm , X^\pm, F$ and $H$. 
Alternatively, one can examine the symplectic form on the space of solutions.
It can be written as 
\be
\omega=\omega_++\omega_-=\frac12\int_{x^-= \infty} dx^+\delta j^-
+ \frac12\int_{x^+= -\infty} dx^-\delta j^+
\>,\label{sf}
\ee
where $\delta$ stands for the exterior derivative on the space of solutions 
and 
$j^{\mu}$ is the symplectic current potential \cite{Witten}. The coefficients
of one half in (\ref{sf}) come from the reflecting boundary 
conditions at $r=0$.  
The light-cone components of the one-form current
$j^{\mu}$ can be easely calculated from the action (\ref{aiii})
\be
j^+=-4\vphi\partial_-\delta\rho-\vphi\partial_-f\delta f-\vphi
e^{-2f}\partial_-h\delta h \>,
\ee
\be
j^-=4\partial_+\vphi\delta\rho-\vphi\partial_+ f\delta f
-\vphi e^{-2f}\partial_+ h\delta h \, .
\ee
By taking into account the asymptotic behaviour of $f$ and $h$ it is easy to 
see that
\bea
\omega&=&\frac12\int_{x^-=-\infty}dx^+\left[\delta X^+\delta\Pi_++\delta F_+
\delta \partial_+F
+\delta H_+\delta \partial_+ H\right]\nonumber\\
&+&\frac12\int_{x^+=\infty}dx^-\left[\delta X^-\delta  \Pi_-+\delta F_-\delta \partial_-F
+\delta H_-\delta \partial_-H\right]
\nonumber\\&=&
\int_{t=const.}dr \left[-\delta X^0\delta \dot X^0 + \delta X^1\delta \dot X^1+\delta F\delta \dot F
+\delta H\delta\dot H\right]
\>,
\eea
where $F_{\pm}(x^{\pm}),H_{\pm}(x^{\pm})$ are the chiral parts of the
 free fields $F$ and $H$,
$(F=F_++F_-,H=H_++H_-)$ and dots represent the $t$ derivatives.

Note that the defining relations for the free fields (\ref{ct1}) and 
(\ref{ct2}) are also valid in the case when $r$ is compact,
and therefore one can have free fields in the case of Gowdy cosmologies.
In that case the corresponding free-field theory is a string theory in 4D
Minkowski space.

Since the cylindrically symmetric gravity can be mapped into a 
set of four free fields with a Minkowskian target space, the quantization in
terms of the free-field variables is considerably simpler than if one uses
the observables obtained from the Ernst equation \cite{KS}, since the later 
lead to a non-linear Yangian algebra. A less straightforward task will be 
finding the
expectation values of the original variables, since they become  complicated
functionals of the free fields. The problem of expressing the original
variables in terms of the observables is in general a complicated problem. 
However, in our case the existence of the free fields $F$ and $H$
implies that one can write an asymptotic series expansions
\be f = {F\over \sqrt\varphi} +{F_1\over\varphi} + 
{F_2\over \varphi\sqrt\varphi} + \cdots \quad, \label{inv1}\ee
and 
\be h = {H\over \sqrt\varphi} +{H_1\over\varphi} + 
{H_2\over \varphi\sqrt\varphi} + \cdots \quad, \label{inv2}\ee
where $F_i$ and $H_i$ are functions of $F$ and $H$, which can be determined
from (\ref{ct2}). In this way one obtains recurrence relations
for higher-order $F_i$ and $H_i$ in terms of the lower order ones, which can be
solved order by order. For example, $F_1 = -\frac12 H\sp 2$ and 
$H_1 = FH$, and so on. When $H =0$, one recovers in this way the asymptotic
expansion of the Bessel function, which is the exact solution in the Abelian
case. Hence the relations (\ref{inv1}) and (\ref{inv2}) can serve as 
explicit expressions for $f$ and $h$ in terms of the free fields $F$ and $H$.
Note that the asymptotic flatness of the dual metric given by (\ref{p2})
requires that $f\rightarrow 0$ and $\tilde h \rightarrow 0$ for 
$r \rightarrow \infty$. Since asymptotically
$\partial_{\pm}\tilde h \sim \mp r dh$ and $h \sim {H\over\sqrt r}$, we then 
obtain that
$H = O (r\sp\e )$, where $\e < -1/2$. 
Note that this asymptotic
behaviour for $H$ corresponds to square integrable functions on the $r$ line.
This is relevant for the quantum case, since this asymptotics gives the Fock
space representations.
 
Note that in the free-field approach the quantum constraints generate a 
2D conformal algebra with a
central charge $c=4$, if the standard quantization of a conformal 
field theory is used.
Consistent quantization can be  then achieved via the introduction of ghost 
fields and background charges in order to have 
vanishing of the total central charge.
Alternatively, if the theory is quantized in the Schrodinger representation, 
the
value of the central charge is $c=2$, because the scalar field with
negative kinetic
energy contributes with $c=-1$ to the Virasoro anomaly\cite{CJZ}.
In order to have a consistent Dirac quantization one has to modify the quantum
constraints in such a way that the anomaly cancels.
The addition to the constraints of a term depending on the pure 2D dilaton 
gravity variables $X^{\pm}$ ensures that
\be
{\tilde C}_{\pm}  =  C_{\pm} + 
{2\over48\pi}\left[{X^{\pm\prime\prime\prime}
\over X^{\pm\prime}}-\left({X^{\pm\prime\prime}\over
X^{\pm\prime}}\right)^2\right]
\>,
\ee
form the constraint  algebra without the anomaly \cite{CJZ}.
The modified constraints can be solved in terms of the 
"gravitationally dressed" oscillators  \cite{Kuchar2}, defined by
\be
\hat F\left(X\right)={1\over2\sqrt{\pi}}\int {dk\over |k|}\left[ e^{ikX}
 \hat a_F\left(k\right)+\ h.c.\right]
\>,
\ee
\be
\hat H\left(X\right)={1\over2\sqrt{\pi}}\int {dk\over |k|}\left[ e^{ikX}
 \hat a_H\left(k\right)+\ h.c.\right]
\>.
\ee

The Fourier coefficients $\hat a_F (k)$ and $\hat a_H (k)$ constitute a complete set 
of observables, so it would be interesting to see how they
are related to the observables obtained from the Ernst equation.
Note that one can construct an $SL(2,R)$ affine algebra 
from the Ernst equation observables, and this algebra generates
the Geroch group \cite{ks2, bj}. On the other hand, one can easily construct
an $SL(2,R)$ affine algebra from $\hat a_F (k)$ and $\hat a_H (k)$ via the
Wakimoto construction \cite{w}. This algebra will be also
a dynamical symmetry algebra. How these two algebras are related would be an
interesting problem for further study.  

We expect that our results can be  extended to the
case of an arbitrary coset space sigma model coupled to 2D dilaton gravity.

\section*{Acknowledgements} 
J. C. acknowledges the Generalitat Valenciana for a F.P.I. fellowship. 
A. M. would like to thank the M.E.C. for a research fellowship.

\end{document}